\documentclass[preprint,3p,times,twocolumn]{elsarticle}
\usepackage{graphicx}
\usepackage{amssymb}
\usepackage{amsmath}
\usepackage{lineno}
\usepackage{booktabs}
\usepackage{float}
\usepackage{comment}
\usepackage{mhchem}
\journal{NIM}
\usepackage{url}
\usepackage[colorlinks,allcolors=red]{hyperref}
\usepackage{upgreek}
\let\affil\address

\newcommand{\Triumph}{1}
\newcommand{\UBC}{2}
\newcommand{\Stanford}{3}
\newcommand{\BNL}{4}
\newcommand{\McGill}{5}
\newcommand{\UMass}{6}
\newcommand{\Erlangen}{7}
\newcommand{\PNL}{8}
\newcommand{\Carleton}{9}
\newcommand{\Duke}{10}
\newcommand{\Illinois}{11}
\newcommand{\ITEP}{12}
\newcommand{\SD}{13}
\newcommand{\LLNL}{14}
\newcommand{\RPI}{15}
\newcommand{\IHEP}{16}
\newcommand{\IME}{17}
\newcommand{\Colorado}{18}
\newcommand{\Sherbrooke}{19}
\newcommand{\Yale}{20}
\newcommand{\Bama}{21}
\newcommand{\Indiana}{22}
\newcommand{\Laurentian}{23}
\newcommand{\Wilmington}{24}
\newcommand{\SLAC}{25}
\newcommand{\Drexel}{26}
\newcommand{\ORNL}{27}
\newcommand{\Stony}{28}
\newcommand{\IBS}{29}

\newcommand{\LHEP}{31}

\affil[\Triumph]{TRIUMF, Vancouver, British Columbia V6T 2A3, Canada}
\affil[\UBC]{Department of Physics and Astronomy, University of British Columbia, Vancouver, British Columbia V6T 1Z1, Canada}
\affil[\Stanford]{Physics Department, Stanford University, Stanford, California 94305, USA}
\affil[\BNL]{Brookhaven National Laboratory, Upton, NY 11973, USA}
\affil[\McGill]{Physics Department, McGill University, Montr{\'e}al, Qu{\'e}bec H3A 2T8, Canada}
\affil[\UMass]{Amherst Center for Fundamental Interactions and Physics Department, University of Massachusetts, Amherst, MA 01003, USA}
\affil[\Erlangen]{Erlangen Centre for Astroparticle Physics (ECAP), Friedrich-Alexander University Erlangen-N\"urnberg, Erlangen 91058, Germany}
\affil[\PNL]{Pacific Northwest National Laboratory, Richland, WA 99352, USA}
\affil[\Carleton]{Department of Physics, Carleton University, Ottawa, Ontario K1S 5B6, Canada}
\affil[\Duke]{Department of Physics, Duke University, and Triangle Universities Nuclear Laboratory (TUNL), Durham, NC 27708, USA}
\affil[\Illinois]{Physics Department, University of Illinois, Urbana-Champaign, IL 61801, USA}
\affil[\ITEP]{Institute for Theoretical and Experimental Physics named by A.~I.~Alikhanov of National Research Center "Kurchatov Institute", Moscow 117218, Russia}
\affil[\SD]{Department of Physics, University of South Dakota, Vermillion, SD 57069, USA}
\affil[\LLNL]{Lawrence Livermore National Laboratory, Livermore, CA 94550, USA}
\affil[\RPI]{Department of Physics, Applied Physics and Astronomy, Rensselaer Polytechnic Institute, Troy, NY 12180, USA}
\affil[\IHEP]{Institute of High Energy Physics, Chinese Academy of Sciences, Beijing 100049, China}
\affil[\IME]{Institute of Microelectronics, Chinese Academy of Sciences, Beijing 100029, China}
\affil[\Colorado]{Physics Department, Colorado State University, Fort Collins, CO 80523, USA}
\affil[\Sherbrooke]{Universit\'e de Sherbrooke, Sherbrooke, Qu{\'e}bec J1K 2R1, Canada}
\affil[\Laurentian]{Department of Physics, Laurentian University, Sudbury, Ontario P3E 2C6, Canada}
\affil[\Wilmington]{Department of Physics and Physical Oceanography, University of North Carolina at Wilmington, Wilmington, NC 28403, USA}
\affil[\Indiana]{Department of Physics and CEEM, Indiana University, Bloomington, IN 47405, USA}
\affil[\Drexel]{Department of Physics, Drexel University, Philadelphia, PA 19104, USA}
\affil[\SLAC]{SLAC National Accelerator Laboratory, Menlo Park, CA 94025, USA}
\affil[\ORNL]{Oak Ridge National Laboratory, Oak Ridge, TN 37831, USA}
\affil[\Bama]{Department of Physics and Astronomy, University of Alabama, Tuscaloosa, AL 35487, USA}
\affil[\Yale]{Wright Laboratory, Department of Physics, Yale University, New Haven, CT 06511, USA}
\affil[\Stony]{Department of Physics and Astronomy, Stony Brook University, SUNY, Stony Brook, NY 11794, USA}
\affil[\IBS]{IBS Center for Underground Physics, Daejeon 34126, Korea}
\affil[\LHEP]{LHEP, Albert Einstein Center, University of Bern, Bern CH-3012, Switzerland}

\begin{document}

\begin{frontmatter}
\title{Characterization of the Hamamatsu VUV4 MPPCs for nEXO}
 \widowpenalty10000
  \clubpenalty10000

\cortext[cor1]{Corresponding author: giacomo@triumf.ca}

\author[\Triumph,\UBC]{G.~Gallina\corref{cor1}}
\author[\Triumph]{P.~Giampa}
\author[\Triumph]{F.~Reti\`{e}re}
\author[\Triumph]{J.~Kroeger\fnref{Heid}}
 \fntext[Heid]{Now at Physikalisches Institut der Universit\"at Heidelberg, Heidelberg, Germany}
 \author[\Triumph]{G.~Zhang\fnref{xian}}
\fntext[xian]{Now at School of Science, Xi'an Polytechnic University, Xi'an, China}
\author[\Triumph]{M.~Ward\fnref{queen}}
\fntext[queen]{Now at Department of Physics, Queen's University, Kingston, Ontario, Canada}
\author[\Triumph]{P.~Margetak}
\author[\Stanford]{G.~Li}
\author[\BNL]{T.~Tsang}
\author[\Triumph]{L.~Doria\fnref{mainz}}
\fntext[mainz]{Now at Institut f\"ur Kernphysik, Johannes Gutenberg-Universit\"at Mainz, Mainz, Germany}
\author[\McGill]{S.~Al Kharusi}
\author[\UMass]{M.~Alfaris}
\author[\Erlangen]{G.~Anton}
\author[\PNL]{I.J.~Arnquist}
\author[\Carleton]{I.~Badhrees\fnref{ARABIA}}
 \fntext[ARABIA]{Also at Home institute King Abdulaziz City for Science and Technology, KACST, Riyadh 11442, Saudi Arabia}
\author[\Duke]{P.S.~Barbeau}
\author[\Illinois]{D.~Beck}
\author[\ITEP]{V.~Belov}%
\author[\SD]{T.~Bhatta}%
\author[\UMass]{J. Blatchford}%
\author[\LLNL]{J.~P.~Brodsky}%
\author[\RPI]{E.~Brown}%
\author[\Triumph,\McGill]{T.~Brunner}%
\author[\IHEP]{G.~F.~Cao\fnref{acascience}}
 \fntext[acascience]{Also at University of Chinese Academy of Sciences, Beijing 100049, China}
\author[\IME]{L.~Cao}%
\author[\IHEP]{W.~R.~Cen}%
\author[\Colorado]{C.~Chambers}%
\author[\Sherbrooke]{S.~A.~Charlebois}%
\author[\BNL]{M.~Chiu}%
\author[\Laurentian]{B.~Cleveland\fnref{snolab}}%
\fntext[snolab]{Also at SNOLAB, Ontario, Canada}
\author[\Illinois]{M.~Coon}%
\author[\Colorado]{A.~Craycraft}%
\author[\Stanford]{J.~Dalmasson}%
\author[\Wilmington]{T.~Daniels}%
\author[\McGill]{L.~Darroch}%
\author[\Indiana]{S.~J.~Daugherty}%
\author[\Triumph,\UBC]{A.~De St. Croix}
\author[\Laurentian]{A.~Der~Mesrobian-Kabakian}%
\author[\Stanford]{R.~DeVoe}%
\author[\Triumph,\UBC]{J.~Dilling}%
\author[\IHEP]{Y.~Y.~Ding}%
\author[\Drexel]{M.~J.~Dolinski}%
\author[\SLAC]{A.~Dragone}%
\author[\Illinois]{J.~Echevers}%
\author[\Carleton]{M.~Elbeltagi}
\author[\ORNL]{L.~Fabris}%
\author[\Colorado]{D.~Fairbank}%
\author[\Colorado]{W.~Fairbank}%
\author[\Laurentian]{J.~Farine}%
\author[\UMass]{S.~Feyzbakhsh}%
\author[\Sherbrooke]{R.~Fontaine}%
\author[\Drexel]{P.~Gautam}
\author[\BNL]{G.~Giacomini}%
\author[\Triumph,\Carleton]{R.~Gornea}%
\author[\Stanford]{G.~Gratta}%
\author[\Drexel]{E.~V.~Hansen}%
\author[\LLNL]{M.~Heffner}%
\author[\PNL]{E.~W.~Hoppe}%
\author[\Erlangen]{J.~H\"{o}{\ss}l}
\author[\LLNL]{A.~House}%
\author[\Bama]{M.~Hughes}%
\author[\McGill]{Y.~Ito\fnref{jaea}}%
\fntext[jaea]{Now at JAEA, Ibaraki, Japan}
\author[\Colorado]{A.~Iverson}%
\author[\Yale]{A.~Jamil}
\author[\Stanford]{M.~J.~Jewell}%
\author[\IHEP]{X.~S.~Jiang}%
\author[\ITEP]{A.~Karelin}%
\author[\Indiana,\SLAC]{L.~J.~Kaufman}%
\author[\UMass]{D.~Kodroff}
\author[\Carleton]{T.~Koffas}%
\author[\Triumph,\UBC]{R.~Kr\"ucken}%
\author[\ITEP]{A.~Kuchenkov}%
\author[\Stony]{K.~S.~Kumar}%
\author[\Triumph,\UBC]{Y.~Lan}%
\author[\SD]{A.~Larson}%
\author[\Stanford]{B.G.~Lenardo}
\author[\IBS]{D.~S.~Leonard}%
\author[\Illinois]{S.~Li}%
\author[\Yale]{Z.~Li}%
\author[\Laurentian]{C.~Licciardi}%
\author[\Drexel]{Y.~H.~Lin}%
\author[\IHEP]{P.~Lv}%
\author[\SD]{R.~MacLellan}%
\author[\McGill]{T.~McElroy}
\author[\McGill]{M.~Medina-Peregrina}
\author[\Erlangen]{T.~Michel}
\author[\SLAC]{B.~Mong}%
\author[\Yale]{D.~C.~Moore}%
\author[\McGill]{K.~Murray}%
\author[\Bama]{P.~Nakarmi}
\author[\ORNL]{R.~J.~Newby}%
\author[\IHEP]{Z.~Ning}%
\author[\Stony]{O.~Njoya}%
\author[\Sherbrooke]{F.~Nolet}%
\author[\Bama]{O.~Nusair}%
\author[\RPI]{K.~Odgers}%
\author[\SLAC]{A.~Odian}%
\author[\SLAC]{M.~Oriunno}%
\author[\PNL]{J.~L.~Orrell}%
\author[\PNL]{G.~S.~Ortega}%
\author[\Bama]{I.~Ostrovskiy}
\author[\PNL]{C.~T.~Overman}%
\author[\Sherbrooke]{S.~Parent}%
\author[\Bama]{A.~Piepke}%
\author[\UMass]{A.~Pocar}%
\author[\Sherbrooke]{J.-F.~Pratte}%
\author[\IME]{D.~Qiu}%
\author[\BNL]{V.~Radeka}%
\author[\BNL]{E.~Raguzin}%
\author[\BNL]{S.~Rescia}%
\author[\Drexel]{M.~Richman}
\author[\Laurentian]{A.~Robinson}%
\author[\Sherbrooke]{T.~Rossignol}%
\author[\SLAC]{P.~C.~Rowson}%
\author[\Sherbrooke]{N.~Roy}%
\author[\PNL]{R.~Saldanha}%
\author[\LLNL]{S.~Sangiorgio}%
\author[\SLAC]{K.~Skarpaas VIII}%
\author[\Bama]{A.~K.~Soma}%
\author[\Sherbrooke]{G.~St-Hilaire}%
\author[\ITEP]{V.~Stekhanov}%
\author[\LLNL]{T.~Stiegler}%
\author[\IHEP]{X.~L.~Sun}%
\author[\UMass]{M.~Tarka}%
\author[\Colorado]{J.~Todd}%
\author[\IHEP]{T.~Tolba\fnref{tb}}%
 \fntext[tb]{Now at the Institut f\"{u}r Kernphysik, Forschungszentrum J\"{u}lich, 52428 J\"{u}lich, Germany}
\author[\McGill]{T.~I.~Totev}%
\author[\PNL]{R.~Tsang}%
\author[\Sherbrooke]{F.~Vachon}%
\author[\Bama]{V.~Veeraraghavan}
\author[\Indiana]{G.~Visser}
\author[\LHEP]{J.-L.~Vuilleumier}
\author[\Erlangen]{M.~Wagenpfeil}
\author[\Laurentian]{M.~Walent}
\author[\IME]{Q.~Wang}
\author[\Carleton]{J.~Watkins}%
\author[\Stanford]{M.~Weber}%
\author[\IHEP]{W.~Wei}%
\author[\IHEP]{L.~J.~Wen}%
\author[\Laurentian]{U.~Wichoski}%
\author[\Stanford]{S.~X.~Wu}%
\author[\IHEP]{W.~H.~Wu}%
\author[\IME]{X.~Wu}
\author[\Yale]{Q.~Xia}%
\author[\IME]{H.~Yang}
\author[\Illinois]{L.~Yang}%
\author[\Drexel]{Y.-R.~Yen}%
\author[\ITEP]{O.~Zeldovich}%
\author[\IHEP]{J.~Zhao}%
\author[\IME]{Y.~Zhou}%
\author[\Erlangen]{T.~Ziegler}

\end{frontmatter}

\section*{Abstract}
In this paper we report on the characterization of the Hamamatsu VUV4 (S/N: S13370-6152) Vacuum Ultra-Violet (VUV) sensitive Silicon Photo-Multipliers (SiPMs) as part of the development of a solution for the detection of liquid xenon scintillation light for the nEXO experiment. Various SiPM features, such as: dark noise, gain, correlated avalanches, direct crosstalk and Photon Detection Efficiency (PDE) were measured in a dedicated setup at TRIUMF. SiPMs were characterized in the range $163 \text{ } \text{K} \leq \text{T}\leq 233 \text{ } \text{K}$. At an over voltage of $3.1\pm0.2$ V and at $\text{T}=163 \text{ }\text{K}$ we report a number of Correlated Avalanches (CAs) per pulse in the $1 \upmu\text{s}$ interval following the trigger pulse of $0.161\pm0.005$. At the same settings the Dark-Noise (DN) rate is $0.137\pm0.002 \text{ Hz/mm}^{2}$. Both the number of CAs and the DN rate are within nEXO specifications. The PDE of the Hamamatsu VUV4 was measured for two different devices at $\text{T}=233 \text{ }\text{K}$ for a mean wavelength of $189\pm7\text{ nm}$. At $3.6\pm0.2$ V and $3.5\pm0.2$ V of over voltage we report a PDE of $13.4\pm2.6\text{ }\%$ and  $11\pm2\%$, corresponding to a saturation PDE of $14.8\pm2.8\text{ }\%$ and $12.2\pm2.3\%$, respectively. Both values are well below the $24\text{ }\%$ saturation PDE advertised by Hamamatsu. More generally, the second device tested  at  $3.5\pm0.2$ V of over voltage is below the nEXO PDE requirement. The first one instead  yields a PDE that is marginally close to meeting the nEXO specifications. This suggests that with modest improvements the Hamamatsu VUV4 MPPCs could be considered as an alternative to the FBK-LF SiPMs for the final design of the nEXO detector.

\section{Introduction}
\label{S:Intro}
Silicon Photo-Multipliers (SiPMs) have emerged as a compelling photo-sensor solution over the course of the last decade \cite{Garutti2011}. In contrast to the widely used Photomultiplier Tubes (PMTs), SiPMs are low-voltage powered, optimal for operation at cryogenic temperatures, and have low radioactivity levels with high gain stability over the time in operational conditions \cite{Baudis2018}. For these reasons, large-scale low-background cryogenic experiments, such as the next-generation Enriched Xenon Observatory experiment (nEXO) \cite{NEXOCollaboration2018}, are migrating to a SiPM-based light detection system \cite{nEXOsipm,Aalseth2018}. nEXO aims to probe the boundaries of the standard model of particle physics by searching for neutrino-less double beta decay (0$\nu \beta \beta$) of  \ce{^{136}Xe} \cite{nEXO_Sensitivity}. 
This lepton number violating process would imply that neutrinos are Majorana fermions.\\The photo-sensor portion of the nEXO experiment must meet the following requirements \cite{NEXOCollaboration2018,Ako}: (i) Photon Detection Efficiency (PDE) greater than 15\% for light at $174.8\pm10.2\text{ nm}$ (scintillation light in liquid xenon \cite{Fujii2015}), (ii) number of correlated avalanches per pulse (within a time window of 1 $\upmu$s after the trigger pulse\footnote{In nEXO, the maximum charge integration time after the trigger will be 1 $\upmu\text{s}$ \cite{NEXOCollaboration2018}}) below 0.2, 
(iii) dark-noise rate lower than $50\text{Hz}/\text{mm}^2$, (iv) electronic noise smaller than 0.1 Photo-electron Equivalent (PE) r.m.s.. Accordingly to recent simulations the first three requirements are in fact sufficient to achieve an energy resolution of $1\%$ for the (0$\nu \beta \beta$) decay of \ce{^{136}Xe} \cite{Ako}. The last requirement is instead a combination of power consumption constraint and total area one channel of the front-end electronics can read out \cite{fabristhesis}. In \cite{Ako}, it was shown that these requirements could be met with SiPMs developed by Fondazione Bruno Kessler (FBK, Trento, Italy). The latest generation of FBK SiPMs significantly exceeds nEXO requirements. The aim of this paper is to assess the performance of Hamamatsu VUV4 Multi-Pixel Photon Counters (MPPC)s  (S/N: S13370-6152) developed for application in liquid xenon as a possible alternative solution to FBK SiPMs. The devices tested have a micro-cell pitch of $50\text{ }\upmu\text{m}$ and an effective photosensitive area of $6\times 6\text{ mm}^2$. Similar characterizations for the same generation of Hamamatsu VUV4 devices, but with different series, can be found in \cite{Baudis2018,Giovanni2018}.

\section{Experimental Details}
\label{S:ExpDet}

\subsection{Hardware Setup}\label{S:hardware}
\begin{figure}[ht]
\centering\includegraphics[width=0.99\linewidth]{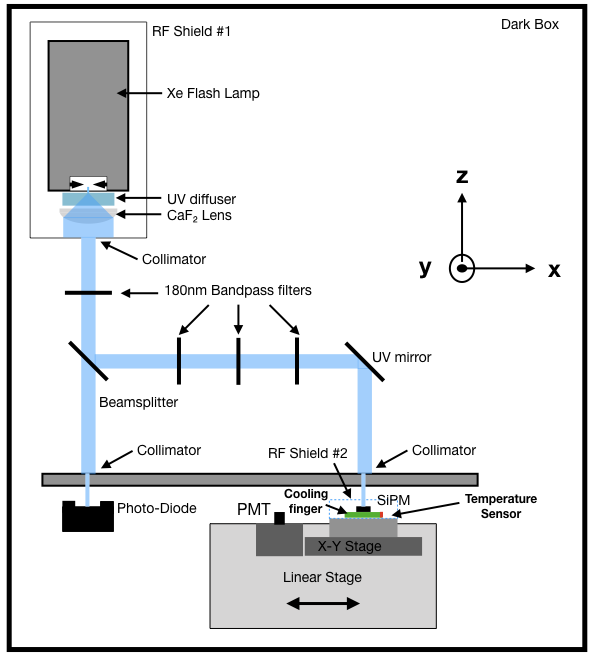}
\centering\includegraphics[width=0.99\linewidth]{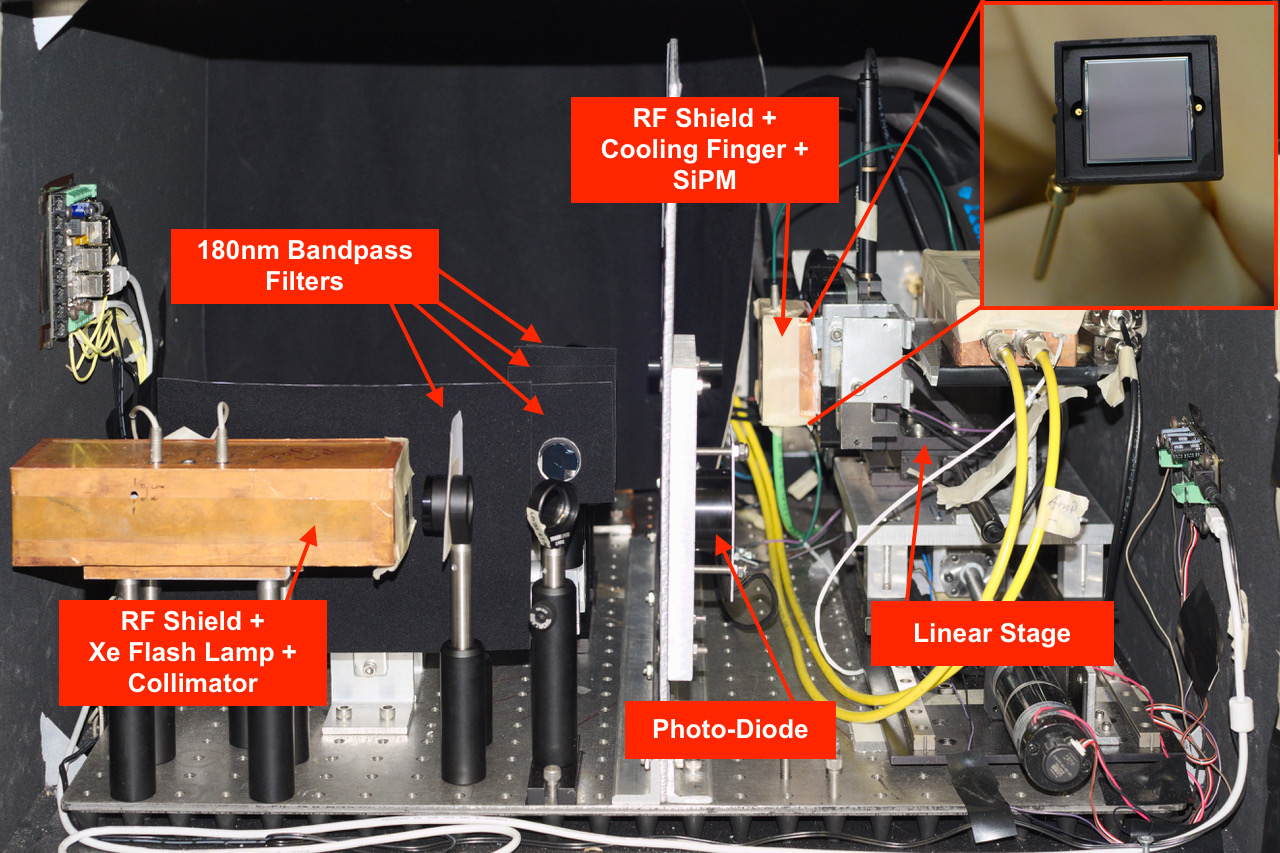}
\caption{Hardware setup used for the characterization of the Hamamatsu VUV4 MPPC.}
\label{F:Setup}
\end{figure}

A setup was developed at TRIUMF to characterize the response of SiPMs at VUV wavelengths, as illustrated in Fig.~\ref{F:Setup}. The measurement setup is installed in a dark box, kept under constant nitrogen gas purge and mounted on an optical platform. Humidity was monitored constantly with a HH314A Omega controller \cite{omega}. Light from a Hamamatsu L11035 xenon flash lamp \cite{pmt} shines through: (i) a 1 mm diameter collimator, (ii) a VUV  diffuser (Knight optical DGF1500 \cite{knigth}) and (iii) a CaF2 lens (eSource optic CF1220LCX \cite{esource}). The diffuser is placed at the focal point of the lens in order to uniformly illuminate the lens and to produce a parallel beam. The whole assembly is housed in a  Radio-Frequency  (RF) copper box (RF Shield \#1 in Fig. \ref{F:Setup}) to shield against electromagnetic emissions from the lamp. Additionally, the VUV light passes through a set of bandpass filters, used to select the desired portion of the lamp emission spectrum and a second 1 mm diameter collimator, close to the SiPM surface (Sec.~\ref{S:filter}). Prior to filters and collimator, the light beam passes through a beam splitter, allowing to simultaneously illuminate the VUV4 MPPC and a Newport 918D-UV-OD3 Photo-Diode (PD) \cite{newport}, used to monitor the xenon flash lamp light stability. The SiPM is mounted inside a second RF shielding copper box (RF Shield \#2 in Fig. \ref{F:Setup}) placed on a movable optical arm which allows for remote positioning of the SiPM in the x-y plane. A liquid nitrogen controlled cold finger adjusts the temperature of the  SiPM down to $163\text{ }\text{K}$ \footnote{The temperature stability was found to be $\pm 1$ K.}. An additional UV-sensitive R9875P Hamamatsu PMT \cite{pmt} is also mounted on the x-y stage and used during PDE measurements to calibrate the absolute light flux at the location of the SiPM. The SiPM and PMT can be interchanged remotely.  A CAEN DT5730B Digitizer Module \cite{caen} and a MIDAS-based control system \cite{MIDAS}, are used for signal digitalization and constitute the Data Acquisition System (DAQ) of the current setup. The SiPM signal is amplified by two MAR6-SM+ amplifiers in series \cite{Amp}. The acquisition trigger module, the PD controller, the filters controller, the x-y stage controller, the PMT and the SiPM power supply are all placed outside the dark box. 

\subsection{Light Filtering Scheme} \label{S:filter}
\begin{figure}[ht]
\centering
\includegraphics[width=0.99\linewidth]{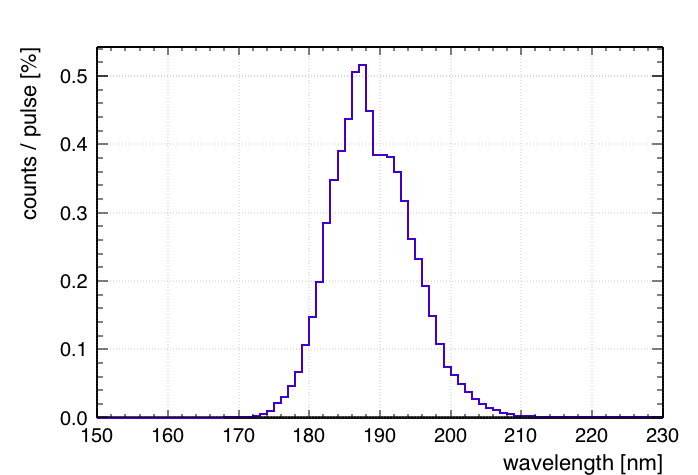}
\caption{Filtered xenon flash lamp spectrum measured after application of the 4 VUV filters. The shoulder effect around 190 nm is due to the combination of the original xenon flash lamp spectrum and the different filters transmissions.}
\label{F:filter_example}
\end{figure}

A set of 4 narrow bandpass filters with transmission nominally centered at 180~nm selects the desired portion of the lamp emission spectrum and attenuates the xenon flash lamp light intensity. Two of the filters are 25180FNB from eSource OPTICS \cite{esource} and two are 180-NB-1D from Pelham Research Optical LLC \cite{pelam}. The typical peak transmission of these filters at $180$ nm is on the order of 20\%. In order to establish the transmitted wavelength distribution, each filter was characterized using a VUV spectrometer (Resonance TR-SES-200 \cite{resonance}). The filtered spectrum is shown in Fig.~\ref{F:filter_example}. It has a mean wavelength of 189\text{ nm} and a Full Width Half Maximum (FWHM) of 13\text{ nm}. 

\begin{figure}[ht]
\centering\includegraphics[width=0.99\linewidth]{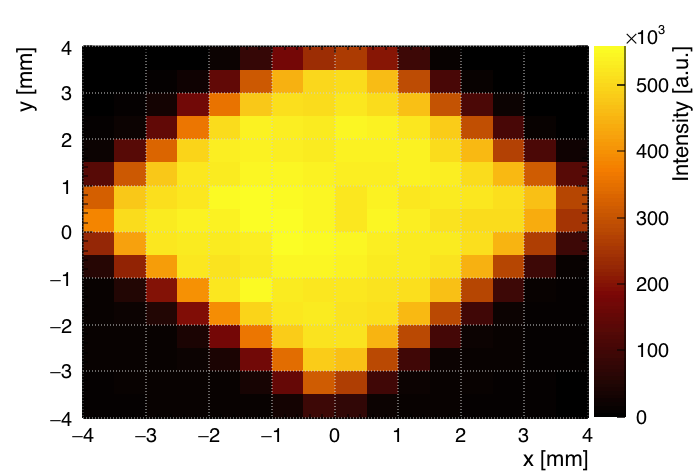}
\caption{Position mapping of the VUV beam at the SiPM location. The SiPM was moved in the x and y coordinates of Fig. \ref{F:Setup} at increments of 0.5 mm in each direction.}
\label{F:beam_map}
\end{figure}

\subsection{Beam Position Mapping}\label{S:beammap}
\begin{figure}[ht]
\centering\includegraphics[width=0.99\linewidth]{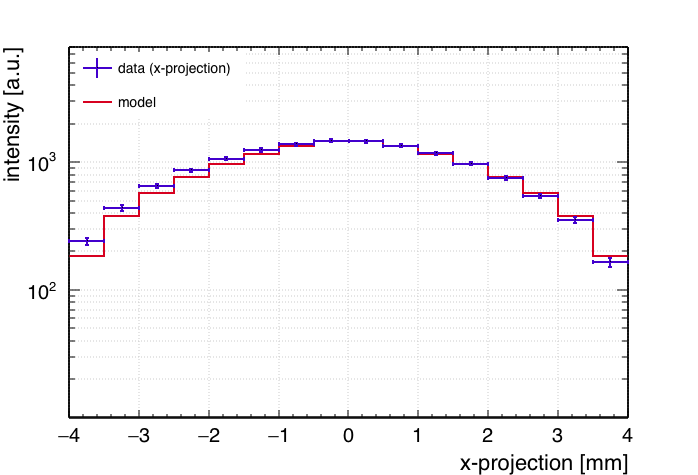}
\centering\includegraphics[width=0.99\linewidth]{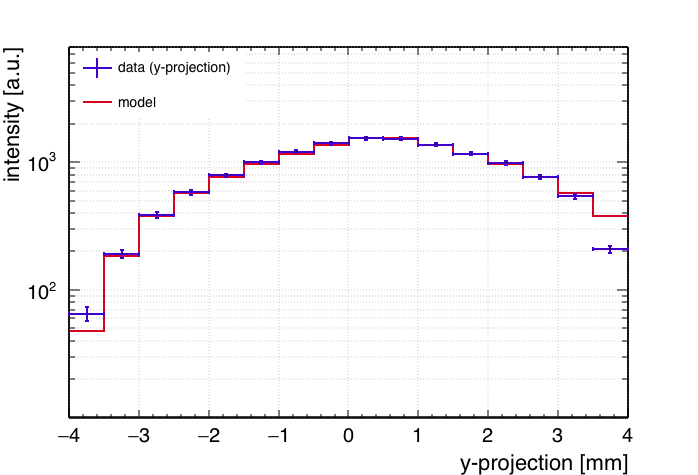}
\caption{Comparison between the measured beam map x-y projections (Blue) and the Monte Carlo model (Red) for the x-projection (Top) and y-projection (Bottom) of the beam map of Fig. \ref{F:beam_map}. The beam diameter was estimated to be  1.2$\pm$0.2 mm at the SiPM and PMT location, indicating that the beam cross-section is smaller than the SiPM and PMT photon-sensitive surface area. The $\chi^2$ test for the two projections gives respectively a $\chi^2/\text{NDF}$ of $1.48$ for the x projection and a  $\chi^2/\text{NDF}$ of $1.24$ for the y projection where NDF is the Number of Degree of Freedom. The biggest deviation between the model and the measurements is evident for points at the edge of the SiPM, where the relative error of the measurement increases due to a worsening of the signal to noise ratio.}
\label{F:beam_fit}
\end{figure}

The VUV4 MPPC and the PMT, used for absolute light calibration, have different Photosensitive Areas (PA) of  $6\times 6 \text{ mm}^2$ and  $50.24\text{  mm}^2$, respectively. For the PDE calculations it is usually necessary to normalize SiPM and PMT measurements by their photosensitive areas. This correction can be neglected if the cross section of the impinging beam is smaller than the SiPM and PMT photosensitive areas. To verify this condition, the collimated light beam was mapped  with respect to the surface of the SiPM, the smaller of the two sensors. This beam map was achieved by moving the SiPM in the x-y plane (Fig.~\ref{F:Setup}) in increments of 0.5 mm, and by measuring (noise subtracted) the counting rates detected by the device at each position. A typical beam-map for the SiPM is shown in Fig.~\ref{F:beam_map}. To size the diameter of the beam: $d_{\text{BEAM}}$, at the SiPM and PMT location, the x and y projections of the beam map are compared with a Monte-Carlo model (MC). Based on the collimator size and shape, described in Sec.~\ref{S:hardware} and Fig. \ref{F:Setup}, the beam map should be the convolution between a circular beam and a square SiPM whose response in the MC model is assumed uniform across its surface. Different x and y profiles were created by letting $d_{\text{BEAM}}$ float, and a likelihood method was used to determine the combination that best fit the data, shown in Fig.~\ref{F:beam_fit}. The best fit  beam diameter at the SiPM (or PMT) location is 1.2$\text{ }\pm\text{ }$0.2 mm. This result confirms that: (i) the beam cross-section is smaller than the SiPM and PMT photon-sensitive areas and therefore an area correction can be neglected during efficiency measurements, (ii) the SiPM response is uniform across its surface. 

\subsection{Collected Data  and Trigger Configurations}
\label{S:DataColl}
The VUV4 MPPC characteristics, as a function of the temperature, were investigated in the range $163 \text{ } \text{K} \leq \text{T}\leq 233 \text{ } \text{K}$. The collected data can be divided into two types: dark data, where the SiPM was shielded from any light source, and xenon flash lamp data. For dark measurements (Sec. \ref{S:gain}, Sec. \ref{S:charge} and Sec. \ref{S:CDP}) the SiPM copper box (RF shield \# 2 in Fig. \ref{F:Setup}) was closed, enabling data-collection at temperatures below $233  \text{ } \text{K}$. The DAQ system was set to trigger on individual SiPM pulses with a DAQ threshold above the noise. For each trigger, the DAQ saves the event with a total sample window of 20 $\upmu$s, split into 4 $\upmu$s of pre-trigger and 16 $\upmu$s of post-trigger samples. The long post-trigger window is necessary in order to measure delayed correlated pulses, while the long pre-trigger window ensures that no tail from previous pulses persists. Dark data were collected at different temperatures and over voltages. For xenon flash lamp data (PDE measurements in Sec \ref{S:PDE}), the DAQ was externally triggered by pulses from a waveform generator which also fired the xenon flash lamp. Furthermore, the SiPM RF shielding box was opened, allowing light to reach the SiPM. PDE measurements were performed at $233  \text{ } \text{K}$ to prevent residual water vapour condensation. The external trigger had  a frequency of 500 Hz. 

\section{Experimental Results} \label{S:ExpRes}

\subsection{Signal Pulse Fitting Procedure}
\label{S:fitsection}
\begin{figure}[ht]
\centering\includegraphics[width=0.99\linewidth]{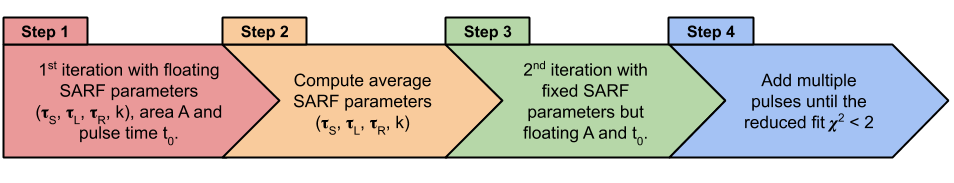}
\caption{Schematic representation of the waveform  analysis  toolkit developed for this paper. See text for a detailed explanation.}
\label{F:FIT_alg}
\end{figure}

The collected dark data were analyzed at the pulse level in order to resolve individual PE pulses. A ROOT-based \cite{ROOT} waveform analysis toolkit was developed, similar to the one reported in \cite{SiPMpulse}. See Fig. \ref{F:FIT_alg}.

\begin{figure}[ht]
\centering\includegraphics[width=0.99\linewidth]{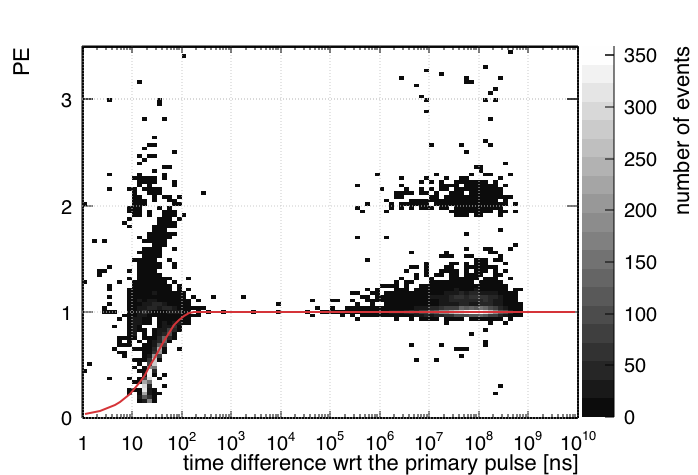}
\caption{Charge distribution (normalized to the average charge of 1 PE pulses) of first pulses following single PE  trigger pulses as a function of the time difference with respect to (wrt) their primary pulse for T=163 K and for an over voltage of $5.1\pm0.2$ V. The gray scale represents the number of events in each bin on a logarithmic scale. The solid red line shows a fit of the afterpulsing events \cite{Otte2016} and is used to measure the recovery time of one cell, equal to $35\pm4\text{ ns}$. For a detailed explanation of the different correlated avalanche mechanisms presented in this figure, we refer the reader to \cite{Ako}.}
\label{F:TNP}
\end{figure}

The implemented algorithm relies on a ${\chi}^{2}$ minimization to identify and fit SiPM pulses. The Single Avalanche Response Function (SARF) was parametrized using the following equation, which accounts for the presence of two time constants in the SiPM pulse shape \cite{FrancescoCorsi2006}:

\begin{equation}
\label{eq:pulseFit}
\begin{aligned}
V(t) = \text{A}\times \Bigg[ &\frac{1-k}{\tau_{S}} \left(e^{-\frac{t-t_{0}}{\tau_{S}+\tau_{R}}} - e^{-\frac{t-t_{0}}{\tau_{R}}}\right) + \\
+ &\frac{k}{\tau_{L}} \left( e^{-\frac{t-t_{0}}{\tau_{L}+\tau_{R}}} - e^{-\frac{t-t_{0}}{\tau_{R}}} \right)\Bigg]
\end{aligned}
\end{equation}
where: $\tau_{R}$ is the pulse rise time, $\text{A}$ is the pulse area (proportional to the pulse charge, see Sec. \ref{S:gain}), $t_{0}$ is the pulse time, $\tau_{S}$ and $\tau_{L}$ are the short and long pulse fall time constants, respectively. $k$ is the relative contribution of the two fall time components in the SiPM pulse shape: $0\leq k \leq 1$. For each over voltage and temperature the collected dark pulses were fitted in a multistage algorithm. First, a pulse-finding algorithm identifies and fits single avalanche pulses to extrapolate the average SARF parameters ($\tau_{R}$,$\tau_{S}$,$\tau_{L}$ and $k$), according to Eq. \ref{eq:pulseFit}. The SiPM pulse shape is then set by fixing these parameters to their estimated average values. Finally a second fit iteration is performed with fixed pulse shape to improve the estimation of pulse time and area. For fits exceeding a certain threshold of reduced ${\chi}^{2}$, test pulses are added iteratively to the fit. The new pulse combination is kept permanently if the reduced ${\chi}^{2}$ of the new fit improves significantly. Else, the test pulses are discarded. The last step of the algorithm improves the capability to identify overlapping pulses (more details in \cite{Ako,Retiere2009}). As an example, we report in Fig. \ref{F:TNP} the charge distribution of first pulses following single PE  primary pulses (also called trigger or prompt pulses) as a function of the time difference with respect to their primary pulse recorded at T=163 K and for an over voltage of $5.1\pm0.2$ V.

\subsection{Single PE Gain and Breakdown Voltage Extrapolation}
\label{S:gain}
\begin{figure}[h]
\centering\includegraphics[width=0.99\linewidth]{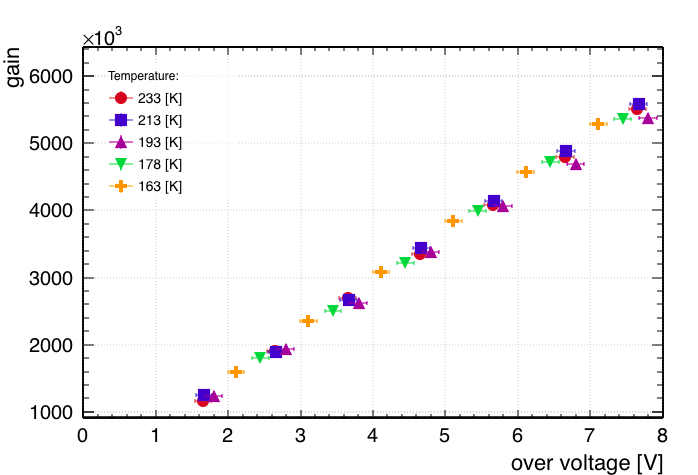}
\caption{Measured single PE gain (Eq.~\ref{eq:gain}) as a function of the over voltage for different  SiPM temperatures.}
\label{F:G_OV}
\end{figure}
Single PE dark pulses were used to measure the single PE SiPM gain as a function of the SiPM temperature. The single PE (amplified) pulse charge $Q$ can be defined as: 

\begin{equation}
Q\equiv \left( \frac{\text{IDR}}{2^{14}} \right) \times \left( \frac{\overline{\text{A}}_{1\text{ PE}} \times 10^{-9}}{\text{R}}\right)
\end{equation}
where: $\text{IDR}$ is the input dynamic range of the digitizer (0.5 V), $\overline{\text{A}}_{1\text{ PE}}$ is the average area (Eq. \ref{eq:pulseFit}) of the fitted digitizer signals for 1 PE pulses (measured in $\text{LSB}\times\text{ns}$ \cite{caen}), and R is the amplifier load resistance (R=50 $ \Omega$). Therefore the single PE (un-amplified) gain $\text{G}_{1\text{ PE}}$, can be extracted from $Q$ using the following equation:

\begin{equation}\label{eq:gain}
\text{G}_{1\text{ PE}}= \frac{Q}{g_{\text{AMP}} \times q_{E}}
\end{equation}
where $g_{\text{AMP}}$ is the gain of the amplifier (measured to be $142\pm2$)\footnote{
The amplifier gain was computed by applying a step voltage to a precision capacitor in order to inject a known charge into the SiPM amplifier input ($i.e.$ $Q_{\text{IN}}$). The output charge was then computed from the digitized amplifier output waveform ($i.e.$ $Q_{\text{OUT}}$). The amplifier gain follows as the ratio between the final integrated charge and the input known charge as  $g_{\text{AMP}}=\frac{Q_{\text{OUT}}}{Q_{\text{IN}}}$.} and $q_{E}$ is the electron charge. In agreement with \cite{Hamamatsu_Manual}, the gain was found to increase linearly with the bias voltage $V$ with a measured gradient of $(72.4\pm3.7)\times {10}^{4}\frac{1}{\text{V}}$. From the single PE gain it is possible to extrapolate the single PE (un-amplified) charge defined as $Q_{1\text{ PE}}\equiv q_E\text{ }\times\text{ }\text{G}_{1\text{ PE}}$. The single PE (un-amplified) charge, as a function of the bias voltage, was then linearly fitted as $Q_{1\text{ PE}}= C_{D}\times\left( V - V_{BD}\right)$, in order to extract: (i) the single SiPM cell capacitance $C_{D}$ and (ii) the Breakdown Voltage $V_{BD}$ defined as the bias voltage at which the SiPM single PE gain (or charge) is zero. The breakdown voltage is found to linearly depend on the SiPM temperature with a measured gradient of $50\pm2\text{ } \frac{\text{mV}}{\text{K}}$. At $163  \text{ } \text{K}$ it is measured to be $V_{BD}=44.9\pm 0.1 $ V. The average cell capacitance, extrapolated from the fit, is $C_{D}=116\pm 6\text{ fF}$ with no observed temperature dependence in the analyzed temperature range. Moreover, the total SiPM capacitance was measured using a capacitance meter (Wayne Kerr 6440B \cite{wayne}), with an extracted single cell capacitance of $C_{D}=104\pm10\text{ fF}$. The gain measurements are shown in Fig.~\ref{F:G_OV} as a function of the over voltage for different SiPM temperatures. 

\subsection{Prompt Pulse Charge Distribution}
\label{S:charge}
\begin{figure}[ht]
\centering\includegraphics[width=0.99\linewidth]{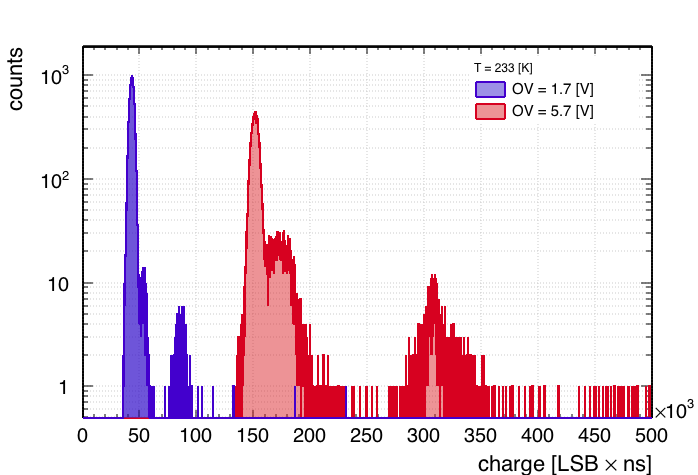}
\centering\includegraphics[width=0.99\linewidth]{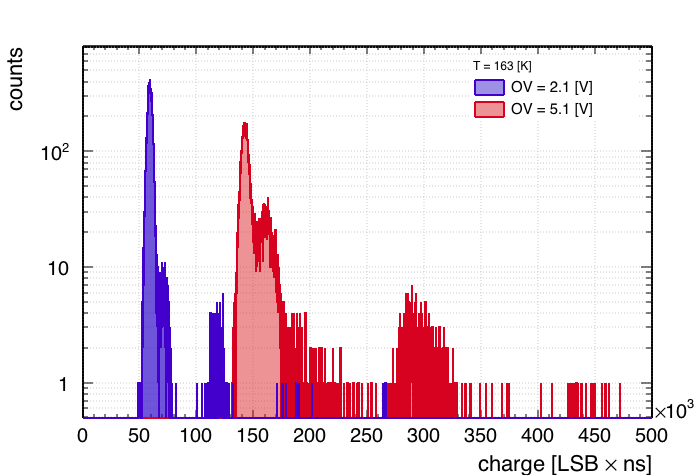}
\caption{Charge distribution for prompt pulses obtained using dark data for different SiPM temperatures and over voltages (Top: T=233 K. Bottom: T=163 K). All the charge peaks show a shoulder-like component to the right of the expected Gaussian charge distribution. This feature appears to be highly dependent on the over voltage.}
\label{F:Amp}
\end{figure}

The pulse-charge distribution was studied using dark data and prompt pulses\footnote{We define a prompt pulse as the first SiPM pulse in each waveform recorded by the DAQ} with the SiPM set at different over voltages and temperatures. The single (and multi) PE charge response of prompt pulses follows the shape of a Gaussian distribution, where the width of the distribution corresponds to the charge resolution combination of the electronic noise and SiPM gain fluctuations. However, we have observed that all the charge peaks show a shoulder-like component to the right of the expected Gaussian charge distribution.
This feature is observed to be highly dependent on over voltage, as shown in Fig.~\ref{F:Amp} for $\text{T}=233  \text{ } \text{K}$ and $163  \text{ } \text{K}$. The shape of the shoulder and its dependence on over voltage were further investigated. We observed that the shape of the pulses which are contributing to the shoulder is consistent with that of single PE with a higher integrated charge. Based on these observations this feature could arise from fast ($<$4 ns) correlated avalanches which do not get resolved from their parent primary pulse by the DAQ system.  This effect could also be explained by a discrepancy in gain between micro-cells in the device.

\subsection{Noise Analysis} 
\label{S:CDP}
Dark and correlated avalanche noise are among the crucial parameters that characterize SiPMs. Dark Noise pulses (DN) are charge signals generated by the formation of electron-hole pairs due to thermionic or field enhanced processes \cite{Hurkx1992}. The DN rate as a function of the applied over voltage for different SiPM temperatures is presented in Sec. \ref{S:CDPSection}. Correlated Avalanche (CA) noise is due to at least two processes: the production of secondary photons during the avalanche in the gain amplification stage detected in nearby cells, and the trapping and subsequent release of charge carriers produced in avalanches. The latter process is usually referred to as afterpulsing, while the former is usually called crosstalk. Crosstalk photons produce nearly simultaneous avalanches to the primary one (Direct CrossTalk (DiCT)) or delayed by several ns (Delayed CrossTalk (DeCT)) \cite{SiPMct}. In general, the subset of the CAs constituted by afterpulses and delayed crosstalk events is named Correlated Delayed Avalanches (CDAs). Unlike DN events, CAs (and therefore CDAs) are correlated with a primary signal. In nEXO the maximum charge integration time after the trigger pulse will be 1 $\upmu \text{s}$. The number of CAs and CDAs per primary pulse in this time window as a function of the applied over voltage for different SiPM temperatures, is reported in Sec. \ref{S:NAAC} and Sec. \ref{S:CDPSection}, respectively. Direct crosstalk is discussed in Sec.~\ref{S:CT}. It is worth recalling that the SiPM noise was studied using dark data, taken with the SiPM shielded from any light source (Sec. \ref{S:DataColl}).

\subsubsection{Number of Correlated Avalanches per Pulse}
\label{S:NAAC}
In order to reach the nEXO design performance, the number of CAs per primary pulse must be below 0.2, within a time window of 1 $\upmu$s after the primary pulse.  A higher value would result in worsened energy resolution \cite{Ako}. 
\begin{figure}[ht]
\centering\includegraphics[width=0.99\linewidth]{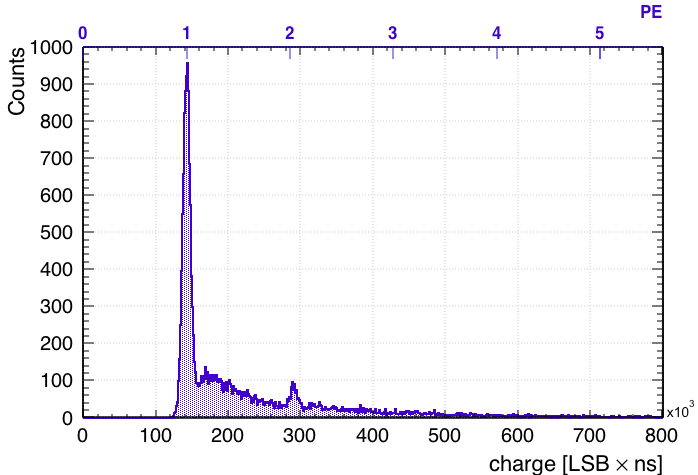}
\caption{Histogram of the baseline subtracted integral of collected dark waveform for an over voltage setting of $5.1\pm0.2$ V and a temperature of 163 K. Each waveform recorded by the DAQ is integrated in the range [0-5]$\upmu\text{s}$. The trigger pulse is at $4\upmu\text{s}$ and no pulses are present in the pre-trigger region ([0-4]$\upmu\text{s}$). The edge next to the 1 PE Gaussian peak is a combination of shoulder-like events (as reported in Fig. \ref{F:Amp}) and waveform with after-pulses.
}
\label{F:MAC1}
\end{figure}
\begin{figure}[ht]
\centering\includegraphics[width=0.99\linewidth]{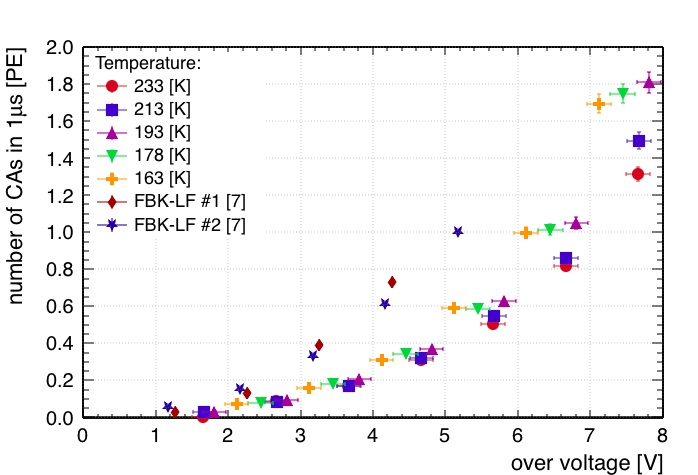}
\caption{Number of Correlated Avalanches (CAs) per primary pulse within a time window of 1 $\upmu$s after the trigger pulse as a function of the applied over voltage for different SiPM temperatures. FBK-LF \#1 and \#2 are instead the number of CAs per pulse, always in a time window of 1 $\upmu$s after the trigger pulse, reported in \cite{Ako}, for two FBK-LF SiPMs.}
\label{F:MAC}
\end{figure}
The average number of CAs per pulse is measured by directly constructing a histogram of the baseline subtracted integral of collected dark waveform in the range $[0-5]\text{ }\upmu\text{s}$ and normalizing it to the average charge of 1 PE pulses, as discussed in \cite{Ako} (An example is reported in Fig. \ref{F:MAC1}). In the absence of correlated noise, the mean of this histogram should be exactly 1. However, each avalanche has a non-zero probability of being accompanied by correlated avalanches causing the mean of the waveform integral histogram to be larger than unity. The excess from unity can be then used to measure (in unit of PE) the average number of CAs per pulse in the 1 $\upmu\text{s}$ interval after the trigger pulse: $\text{N}_{\text{CA}}$. The $\text{N}_{\text{CA}}$ number is reported in Fig. \ref{F:MAC}, for different SiPM temperatures and over voltages. At low over voltages $\text{N}_{\text{CA}}$ doesn't show an evident temperature dependence; a minor deviation is, however, observed at higher over voltages. Similar trends are reported for other SiPMs by other investigations \cite{SiPMFBKDS20k}. In order to meet the nEXO requirements (Sec. \ref{S:Intro}), the Hamamatsu VUV4 can be operated at T=$163  \text{ } \text{K}$ up to $4$ V of over voltage keeping $\text{N}_{\text{CA}}$ smaller than 0.2. For comparison, in Fig. \ref{F:MAC}, we have also reported the average number of CAs per pulse of another SiPM produced specifically for nEXO by FBK, the FBK Low-Field (LF) SiPM (FBK-LF \#1 and FBK-LF \#2 in Fig. \ref{F:MAC}), characterized in a previously reported work \cite{Ako}. The Hamamatsu VUV4 has a considerably lower $\text{N}_{\text{CA}}$ than FBK-LF. It can therefore be operated at higher OV (i.e. higher gain) while meeting nEXO requirements ($\text{N}_{\text{CA}}\leq 0.2$). For reference, at T=$163  \text{ } \text{K}$ and at $3.1\pm0.2$ V of over voltage we measure $\text{N}_{\text{CA}}=0.161\pm0.005$ for the Hamamatsu VUV4\footnote{$3.1\pm0.2$ V is the closest measured points for which the requirement on $\text{N}_{\text{CA}}$ is satisfied. At this over voltage and temperature the electronics noise was measured to be 0.064 PE.}, while the FBK LF \#1 and \#2 at $\text{T}=168\text{ } \text{K}$ and $3.26$ V and $3.17$ V have a $\text{N}_{\text{CA}}$ equal to $0.39$ and $0.33$, respectively.

\subsubsection{Dark Noise Rate and Number of Correlated Delayed Avalanches per Pulse}
\label{S:CDPSection}
\begin{figure}[ht]
\centering\includegraphics[width=0.99\linewidth]{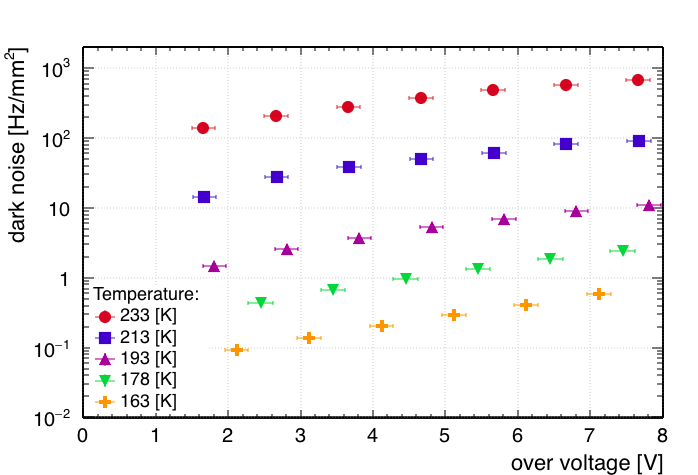}
\caption{Dark Noise (DN) rate normalized by the SiPM photon sensitive area as a function of the applied over voltage for different SiPM temperatures.}
\label{F:DN_OV}
\end{figure}

\begin{figure}[ht]
\centering\includegraphics[width=0.99\linewidth]{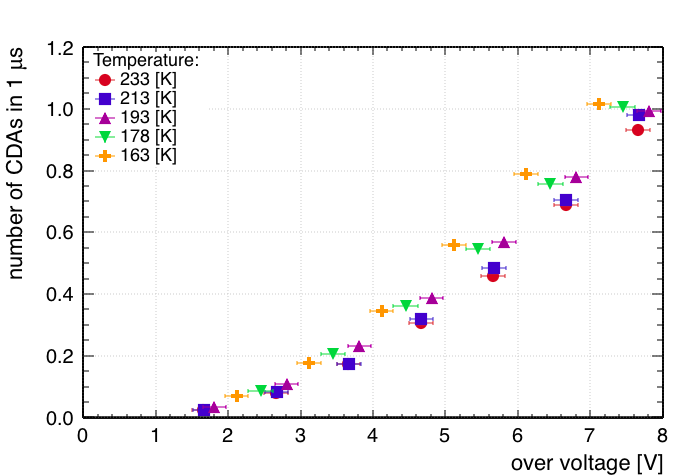}
\caption{Number of Correlated Delayed Avalanches (CDAs) per primary pulse within a time window of 1 $\upmu$s after the trigger pulse as a function of the applied over voltage for different SiPM temperatures.}
\label{F:AP_OV}
\end{figure}

DN and CDAs events can be distinguished by studying the time distribution of events relative to the primary pulse using a method, described in \cite{SiPMcdp}, that requires the charge of the primary pulse to be a single PE equivalent. The observed pulse rate $\text{R}(t)$ is then computed as a function of the time difference $t$ from the primary pulse ($t=0$) as:

\begin{equation}\label{eq:rt}
    \text{R}(t) = \text{R}_{\text{DN}}(t)+ \text{R}_{\text{CDA}}(t)
\end{equation}
where $\text{R}_{\text{DN}}$ is the rate of dark noise pulses and $\text{R}_{\text{CDA}}$ is the rate of the CDAs per pulse. 
\begin{figure}[ht]
\centering\includegraphics[width=0.99\linewidth]{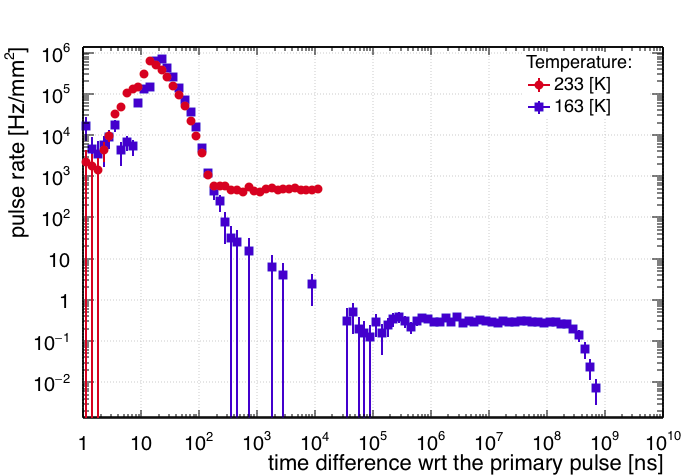}
\caption{Observed pulse rate $\text{R}(t)$ normalized by the SiPM photon sensitive area as a function of the time difference with respect to (wrt) the primary pulse for a temperature of $163  \text{ } \text{K}$ ($5.1\pm0.2$ V of over voltage) (Blue) and $233  \text{ } \text{K}$ ($5.7\pm0.2$ V of over voltage) (Red). At $233  \text{ } \text{K}$, to avoid ADC dead time corrections due to the high self trigger DN rate, the time differences were computed only between pulses inside the same DAQ window (maximum time difference: $16\upmu$s)}
\label{F:AP_TTNP}
\end{figure}
An example of the measured time distribution for a given over voltage and for two different temperatures is reported in Fig.~\ref{F:AP_TTNP}. The DN rate can then be estimated from Fig.~\ref{F:AP_TTNP} performing a weighted mean of the asymptotic rate at long times\footnote{Each point in the mean was weighted with its error derived as reported in \cite{SiPMcdp}.}. The measured DN rates, for different SiPM temperatures, are shown in Fig.~\ref{F:DN_OV} as a function of the over voltage. For the same temperature and OV setting reported in Sec. \ref{S:NAAC} ($163  \text{ } \text{K}$, $3.1\pm0.2$ V of OV),  the DN rate is measured to be  $0.137\pm0.002 \text{ Hz/mm}^{2}$, which satisfies nEXO requirements (Sec. \ref{S:Intro}). By applying Eq. \ref{eq:rt} to the observed pulse rate (e.g. Fig.~\ref{F:AP_TTNP}), it is possible to compute the expected number of CDAs per pulse in a fixed time window of length $\Delta t$ after the trigger pulse as: 

\begin{equation}
\text{N}_{\text{CDA}}(\Delta t) = \int_{0}^{\Delta t} \Big( \text{R}(t)- \text{R}_{\text{DN}}(t) \Big)~\text{dt} 
\end{equation}
where $\text{R}_{\text{DN}}$ is the measured DN rate as reported in Fig.~\ref{F:DN_OV}. The measured average number of CDAs per pulse in the 1 $\upmu \text{s}$ window after the trigger pulse, for different SiPM temperatures and over voltages is reported in Fig.~\ref{F:AP_OV}. For $163  \text{ } \text{K}$ and $3.1\pm0.2$ V of OV, the average number of CDAs per pulse in 1 $\upmu$s is $0.178\pm0.003$. Finally, it is worth noting that the number of CDAs, extrapolated in this section, cannot be compared directly with the number of CAs (Sec. \ref{S:NAAC}). The number of CDAs is in fact derived, accordingly to \cite{SiPMcdp}, considering only the time differences of delayed events with respect to their primary pulse while the number of CAs accounts instead also for their different charge.

\subsubsection{Number of Additional Prompt Avalanches} \label{S:CT}
\begin{figure}[ht]
\centering\includegraphics[width=0.99\linewidth]{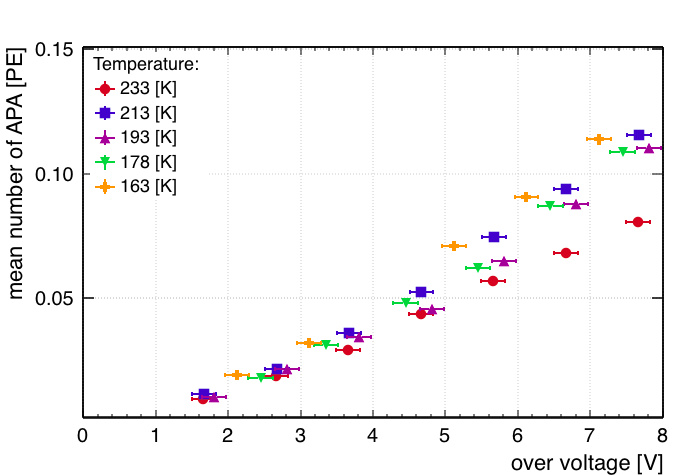}
\caption{Number of Additional Prompt Avalanches (APAs) as a function of the over voltage for different SiPM temperatures.}
\label{F:CT_OV} 
\end{figure}

Based on the measured dark noise rate reported in Sec.~\ref{S:CDPSection}, and assuming Poisson statistics, the probability of having two dark noise pulses occurring within few nano-seconds is negligible. Therefore, the collected dark data can be used to investigate Direct CrossTalk (DiCT). DiCT occurs when photons generated during a triggered avalanche in one micro-cell promptly travel to the amplification region of neighboring micro-cells, where they induce a secondary avalanche. This mechanism happens over pico-seconds \cite{SiPMct} and it mimics a multiple PE signal, thus biasing the photon counting ability of the device. The charge distribution of the prompt pulses obtained from the dark data (Sec. \ref{S:charge}) can be used to determine the mean number of Additional Prompt Avalanches (APA)s, $\text{N}_{\text{APA}}$, due to Direct CrossTalk :

\begin{equation}
\text{N}_{\text{APA}}=\frac{1}{N}\sum_{i=1}^{N}\frac{\text{A}_i}{\overline{\text{A}}_{\text{1 PE}}}-1
\end{equation}
where $\text{A}_i$ is the charge of the prompt pulse $i$ (Sec. \ref{S:fitsection}), $\overline{\text{A}}_{\text{1 PE}}$ is the average charge of 1 PE pulses and $N$ is the number of prompt avalanches analyzed. The $\text{N}_{\text{APA}}$ number, in unit of PE, as a function of the over voltage and for different SiPM temperatures, is reported in Fig.~\ref{F:CT_OV}. For $163  \text{ } \text{K}$ and for $3.1\text{ }\pm\text{ }0.2$ V of OV we measure a mean number of additional prompt avalanches, due to direct cross talk, equal to $0.032\pm 0.001$. The small amount of $\text{N}_{\text{APA}}$ can also be understood looking at Fig.~\ref{F:TNP} and Fig.~\ref{F:MAC1} where 3 PE events are greatly suppressed.

\subsection{Photon Detection Efficiency}\label{S:PDE}
\begin{figure}[ht]
\centering\includegraphics[width=0.99\linewidth]{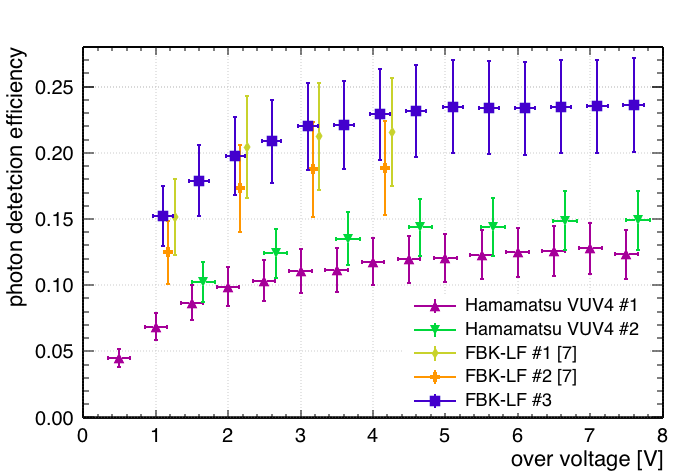}
\caption{Photon Detection Efficiency (PDE) as a function of the over voltage for two Hamamatsu VUV4 MPPCs and three FBK-LF SiPMs. The Hamamatsu MPPC used throughout this paper is the VUV4 \#2. The FBK-LF \#3 is a new FBK LF SiPM measured, for comparison, in the setup reported in this paper. FBK-LF \#1 and \#2 are instead the FBK-LF PDE reported in \cite{Ako}, whose number of CAs is reported in Sec. \ref{S:NAAC} and Fig. \ref{F:MAC}. The error, on each point, for all the five devices accounts for the presence both to the statistical and the systematic uncertainty.}
\label{F:PDE}
\end{figure}

To meet nEXO requirements, the SiPM PDE  must be $\geq15\%$ \cite{Ako} for liquid xenon scintillation wavelengths ($174.8\pm10.2\text{ nm}$ \cite{Fujii2015}). In this paper, the PDE was measured using a filtered pulsed xenon flash lamp enabling measurements free from correlated avalanches \cite{Otte2016}. As shown in Sec.~\ref{S:filter}, the mean wavelength after filtering is $189\pm7\text{ nm}$. The generated flash pulse has a width $\sim1\upmu$s, which prevents room temperature measurements, as the dark noise would overwhelm the light signal. For this reason PDE data were taken with the SiPM temperature set at 233 K. The PDE was measured using an experimental technique similar to the one described in \cite{Otte2016}. The light flux was assessed using a calibrated PMT (Sec. \ref{S:hardware}) with no corrections applied for the different photosensitive areas of the two devices (Sec. \ref{S:beammap}). The average number of photons detected by the SiPM ($\upmu_{\gamma}^{\text{SiPM}}$) was measured by counting the number of lamp flashes in which no pulses were detected ($N_{0}$). Using Poisson statistics $\upmu_{\gamma}^{\text{SiPM}}$ can be expressed as:

\begin{equation}
\upmu_{\gamma}^{\text{SiPM}} = -\ln\left( \frac{N_{0}}{N_{\text{TOT}}}\right) - \upmu_{\text{DN}}
\end{equation}
where $N_{\text{TOT}}$ is the total number of lamp flashes. This method is independent of correlated avalanches and it  requires only correcting for the average number of dark noise pulses in the trigger window ($\upmu_{\text{DN}}$). The same analysis was applied to the reference PMT (Hamamatsu R9875P, Sec. \ref{S:hardware}) in order to  measure the average number of photons detected by the PMT  ($\upmu_{\gamma}^{\text{PMT}}$) and then extrapolate, using the PMT Quantum Efficiency (QE) and Collection Efficiency (CE), the effective number of photons produced by the pulsed xenon flash lamp at the SiPM and PMT surface: $N_{\gamma}$. The PMT QE expresses the probability of photon-electron emission when a single photon strikes the PMT photocathode \cite{pmtbook}. The QE reported by Hamamatsu for this phototube is  $16.5\pm2.1$ \%. The PMT CE expresses the probability that the generated photon-electron will land on the effective area of the first PMT dynode and start the amplification process \cite{pmtbook}. The CE of this phototube was determined by Hamamatsu to be $71$\% without any estimate of the uncertainty. Conservatively, in agreement with \cite{Ako}, we assume a $10$\% error for the CE of this PMT (i.e. $71\pm10$\%) to account for the non-uniformity of the photon collection at the photo-cathode. The number of incident photons $N_{\gamma}$ on the PMT surface can then be obtained as:

\begin{equation}
N_{\gamma} = \left( \frac{\upmu_{\gamma}^{\text{PMT}}}{\text{CE}\times\text{QE}} \right)
\label{pmtngamma}
\end{equation}
The SiPM PDE is in turn obtained as:

\begin{equation}
\text{PDE}_{\text{SiPM}} = \left( \frac{\upmu_{\gamma}^{\text{SiPM}}}{N_{\gamma}} \right)
\end{equation}
The measured PDE as a function of the over voltage is shown in Fig.~\ref{F:PDE} for two different VUV4 devices labelled VUV4 \#1 and VUV4 \#2. VUV4 \#2 is the device for which DN, CAs, CDAs and APAs are reported in the previous sections of this paper. The measured saturation PDE for VUV4 \#2 and VUV4 \#1 are $14.8\pm2.8\%$ and $12.2\pm2.3\%$, respectively. For comparison, we measured the PDE of one FBK LF (FBK LF \# 3 in Fig. \ref{F:PDE}), for which the saturation PDE was measured to be $22.8\pm4.3\%$, in agreement with \cite{Ako} (FBK-LF \#1 and \#2 in Fig. \ref{F:PDE}, see Sec. \ref{S:NAAC}). It is important to mention that different sources of systematic uncertainty were considered and investigated for this measurement. The PMT gain stability was found to be a negligible source of uncertainty. The stability of the light flux, monitored with a photo-diode (Sec.~\ref{S:hardware}), was also found to have a negligible effect, with fluctuations within 1 \%. The dominant source of systematic uncertainty is therefore the uncertainty on the PMT CE. \\ We conclude this section by interpreting the results of the PDE measurement in the context of the nEXO experiment. There are three main differences between the experimental condition of this study and the nEXO environment: (i) wavelength, (ii) angle of incidence of the light, and (iii) temperature. Firstly, as shown in Sec. \ref{S:filter}, the filtered wavelength distribution presented in this work is about 10 nanometers higher than liquid xenon scintillation. In \cite{Doc_VUV4} Hamamatsu reports little wavelength dependence in this range. Secondly, as shown in Sec. \ref{S:beammap}, the angular distribution of photons at the SiPM location in the TRIUMF setup is normal to the SiPM surface. This is not the case for nEXO, for which preliminary GEANT4 optical simulations \cite{geant} show an average deviation from a normal distribution at the SiPM surface. The photon detection efficiency is likely to be lower at higher incidence angles due to increased reflectivity \cite{Otte2016}. Finally, as explained in Sec. \ref{S:ExpDet}, the VUV4 PDE is measured at $233\text{ } \text{K}$ instead of liquid xenon temperature, which may affect the total PDE \cite{Collazuol2011,Johnson2009}. Measurements of SiPM reflectivity to VUV light as a function of photon incident angle and wavelength are being planned with setups properly customized to perform PDE measurements down to $163\text{ } \text{K}$. These custom setups will allow us to measure the VUV4 PDE in conditions closer to those in nEXO. A lower systematic uncertainty on the VUV4 PDE will be also achieved substituting the calibrated Hamamatsu PMT with a calibrated NIST photodiode considering the smaller systematic uncertainty on the provided responsivity in the VUV range \cite{NIST} if compared with the PMT CE.

\section{Conclusions}

\begin{table}
\centering

\begin{tabular}{ccc}
   \toprule
   Quantity & Value & Unit \\
   \midrule
   Dark noise rate & $0.137\pm0.002$  & Hz/mm$^{2}$\\
   Number of CAs  & $0.161\pm0.005$ & PE\\
   Number of APAs & $0.032\pm 0.001$ & PE \\
   Number of CDAs & $0.178\pm0.003$ & - \\
   PDE VUV4 \#2 & $13.4\pm2.6\text{ }$\%  & -\\
   PDE VUV4 \#1 & $11\pm2\text{ }$\% & - \\
   \bottomrule
\end{tabular} 
\caption{Summary of the results derived for the characterization of the Hamamatsu VUV4 MPPC useful for nEXO operation. The dark noise rate, the number of Correlated Avalanches (CAs) per pulse in the $1\upmu\text{s}$ following the trigger pulse, the number of Additional Prompt Avalanches (APAs) and the number of Correlated Delayed Avalanches (CDAs) per pulse in the same time window are reported for a temperature of $163  \text{ } \text{K}$ and at $3.1\pm0.2$ V of Over Voltage (OV). The Photon Detection Efficiency (PDE) was measured for two Hamamatsu VUV4 devices (labelled as VUV4 \#2 and VUV4 \#1) at $\text{T}=233 \text{ }\text{K}$, for a mean wavelength of $189\pm7\text{ nm}$ and at an over voltage of $3.6\pm0.2$ V and $3.5\pm0.2$ V, respectively.
}
\label{T:1}
\end{table}
This paper describes measurements performed at TRIUMF to characterize the properties of VUV sensitive SiPMs at cryogenic temperatures. In particular, this work focused on the Hamamatsu VUV4 MPPC, identified as a possible option for the nEXO experiment. The results of the characterization are summarized in Table \ref{T:1}. For a device temperature of $163  \text{ } \text{K}$, the VUV4 dark noise rate is $0.137\pm0.002 \text{ Hz/mm}^{2}$ at $3.1\pm0.2$ V of over voltage, a level comfortably lower than what required for nEXO ($<$50 Hz/mm$^{2}$). At the same over voltage setting and temperature, we measure: a number of additional prompt avalanches equal to $0.032\pm 0.001$, a number of correlated delayed avalanches per pulse in the $1\upmu\text{s}$ following the trigger pulse of  $0.178\pm0.003$ and a number of correlated avalanches per pulse in the same time window equal to $0.161\pm0.005$, also consistent with nEXO requirements. Finally, the PDE of the Hamamatsu VUV4 was measured for two different devices (labelled as VUV4 \#2 and VUV4 \#1) at $\text{T}=233 \text{ }\text{K}$. At $3.6\pm0.2$ V and $3.5\pm0.2$ V of over voltage we measure, for a mean wavelength of $189\pm7\text{ nm}$, a PDE of $13.4\pm2.6\text{ }\%$ and  $11\pm2\%$ for the two devices, corresponding to a saturation PDE of $14.8\pm2.8\text{ }\%$ and $12.2\pm2.3\%$, respectively. Both values are well below the $24\text{ }\%$ saturation PDE advertised by Hamamatsu \cite{Doc_VUV4}\footnote{This discrepancy could be related to the different technique used by Hamamatsu to evaluate the VUV4 PDE. Accordingly to an internal meeting with Hamamatsu the PDE reported in \cite{Doc_VUV4} was not assessed by pulse counting, but reading out instead the MPPC current under illumination. However, as shown in \cite{Girard2018}, the MPPC current is affected by CA noise and it is therefore easier to overestimate the PDE if the CA noise contribution is not accounted properly.}. More generally, the VUV4 \#1 at 3.5 V of over voltage is below the nEXO PDE requirement. The VUV4 \#2 instead yields a PDE that is marginally close to meeting the nEXO specifications. This  suggests that with modest improvements the Hamamatsu VUV4 MPPCs could be considered as an alternative to the FBK-LF SiPMs for the design of the nEXO detector. 

\section{Acknowledgments}
This work has been supported by NSERC, CFI, FRQNT, NRC, and the McDonald Institute (CFREF) in Canada;  by the Offices of Nuclear and High Energy Physics within DOE's Office of Science, and NSF in the United States; by SNF in Switzerland; by IBS in Korea; by RFBR (18-02-00550) in Russia; and by CAS and ISTCP in China. This work was supported in part by Laboratory Directed Research and Development (LDRD) programs at Brookhaven National Laboratory (BNL), Lawrence Livermore National Laboratory (LLNL), Oak Ridge National Laboratory (ORNL), and Pacific Northwest National Laboratory (PNNL).

\section{References}

\bibliographystyle{unsrt}
\bibliographystyle{abbrvnat}
\bibliography{sample}


\end{document}